\documentclass[aps,twocolumn,noshowpacs,superscriptaddress,floatfix,nofootinbib]{revtex4-1}
\usepackage[table]{xcolor}
\usepackage{quantikz}
\usepackage{amsmath,amssymb}
\usepackage{mathrsfs}
\usepackage[justification=justified]{caption}
\usepackage{xcolor}
\usepackage{verbatim}
\usepackage[normalem]{ulem}
\usepackage[rightcaption]{sidecap}
\usepackage{graphicx}
\usepackage{subfigure}
\usepackage{caption}
\usepackage{multirow}
\usepackage{amsthm,epsfig,tikz}
\usepackage{float}
\usepackage{multirow}
\usepackage[colorlinks,citecolor=black,urlcolor=black,linkcolor =black,bookmarks=false,hypertexnames=true]{hyperref} 
\usepackage{relsize}

\usepackage{float}
\makeatletter
\let\newfloat\newfloat@ltx
\makeatother
\usepackage{algorithm}
\usepackage{algcompatible}

\renewcommand{\part}[2]{\frac{\partial #1}{\partial #2}}

\DeclareMathOperator*{\argmin}{arg\,min}

\usepackage{float}
\makeatletter
\let\newfloat\newfloat@ltx
\makeatother

\begin{document}

\title{An Advantage Using Feature Selection with a Quantum Annealer}

\author{Andrew Vlasic}
\email{avlasic@deloitte.com}
\affiliation{Deloitte Consulting, LLP}

\author{Hunter Grant}
\affiliation{Deloitte Consulting, LLP}

\author{Salvatore Certo}
\affiliation{Deloitte Consulting, LLP}

\date{\today}

\begin{abstract}
Feature selection is a technique in statistical prediction modeling that identifies features in a record with a strong statistical connection to the target variable. Excluding features with a weak statistical connection to the target variable in training not only drops the dimension of the data, which decreases the time complexity of the algorithm, it also decreases noise within the data which assists in avoiding overfitting. In all, feature selection assists in training a robust statistical model that performs well and is stable. Given the lack of scalability in classical computation, current techniques only consider the predictive power of the feature and not redundancy between the features themselves. Recent advancements in feature selection that leverages quantum annealing (QA) gives a scalable technique that aims to maximize the predictive power of the features while minimizing redundancy. As a consequence, it is expected that this algorithm would assist in the bias/variance trade-off yielding better features for training a statistical model. This paper tests this intuition against classical methods by utilizing open-source data sets and evaluate the efficacy of each trained statistical model well-known prediction algorithms. The numerical results display an advantage utilizing the features selected from the algorithm that leveraged QA.
\end{abstract}

\maketitle

\section{Introduction}\label{sect:intro} 
This manuscript gives a detailed numerical analysis of the efficacy of the quantum feature selection algorithm \cite{mucke2022quantum} (QFS) against the features selected from the ANOVA F-test \cite{dhanya2020f} and the features selected from the chi-squared statistic \cite{jin2015chi}. Efficacy determined through the evaluation of the statistical models trained from each set of selected features. Three different data sets are applied and the outputs from each feature selection algorithm are then fed into standard algorithms that train a statistical prediction model. The evaluations of the statistical models are then aggregated and analyzed. Consistently it is shown that the features selected by QFS yield higher performing and stable statistical models. 

Recent advancements in classical computing has allowed for an exponential growth in statistical predictive modeling within academia as well as industry. The increased computational capacity of classical computers has also proliferated the collection data, resulting in a vast sets of data that is rich is information. However, the incorrect utilization of this information has the potential to add noise to the training of a statistical model. For instance, take the task of statistically modeling fraud \cite{belhadji2000model}. Within this paper it was shown that only half of the features were useful in predicting fraud. In all, the current size and dimensionality of information has necessitated the need for analyzing the data to decrease noise and collinearity of features \cite{dormann2013collinearity, gupta2019dealing}. 

Two well-applied feature selection methods are the ANOVA F-test and chi-squared statistics. While these two statistical measures have been well-used, the methods only consider the statistical strength between the features and the target variable. In particular, while these methods can decrease noise in the independent variables, they do not decrease redundant information, which has the potential to create a biased statistical model since the algorithm is only given relevant information to derive a model; see \cite{li2008method, kalnins2018multicollinearity, danasingh2020identifying, zhao2019maximum, grillo2021redundancy} for a comprehensive view.  

The feature selection algorithm QFS is a \textit{quadratic unconstrained binary optimization} (QUBO) problem that finds the appropriate weight for the linear terms, which is the signal between a feature and the target variable, and the appropriate weight for the quadratic terms, the shared information between each feature. This algorithm separates the signal from the noise by identifying the features with a strong statistical connection with the target variable while not including other features that, while having similarly strong statistical connections with the target variable, also have redundant information that is already captured by other features.

For background, a QUBO is an optimization problem of the form 
\begin{equation}
    \mbox{minimize/maximize} \ x^t Q x
\end{equation}
where $x$ is a vector of binary decision variables and $Q$ is a constant square matrix. Section \ref{sect:qa} describes the connection of a QUBO optimization problem to physics. Please see \cite{glover2018tutorial} for further information. 

While the QUBO described in the paper \cite{mucke2022quantum} can be solved with classical computational methods with a small number of features, as the number of features increase, which increases the quadratic terms in a factorial manner, it becomes progressively difficult for these methods to find solutions in a reasonable time.  Unlike many other problem formulations in operations research, this application is a natural quadratic programming problem and quantum annealers (QA) are natural tools to use since QA scales well as the number of binary variables grow \cite{albash2018adiabatic,albash2018demonstration}. For the application of feature selection see \cite{nembrini2021feature, von2021quantum}.

\section{Applications of Quantum Annealing} \label{sect:qa}
Quantum computers are theorized to solve problems currently intractable on classical computers, potentially offering exponential speedups to some of the hardest problems. Quantum annealers are processors specifically designed to solve combinatorial optimization problems, of which there are exciting applications in similar areas, especially in areas like unsupervised learning and training classical AI models \cite{schuman2019classical, neven2012qboost}.


Quantum annealing follows the adiabatic theorem, which states that if the time evolution is long enough, a quantum system will stay in the ground state of its Hamiltonian. As described in the foundation work in \cite{farhi2000quantum}, for a quantum system starting in a ground state of a Hamiltonian $H_b$ and evolving to the ground state of the problem Hamiltonian $H_p$, the system can be described mathematically at time $t$ out of the total run time $T$ as
\begin{equation*}
\mathcal{H}(t) = \big( 1-u(t) \big)\mathcal{H}_b + u(t)\mathcal{H}_p,
\end{equation*}
where $u(t) = t/T$. 

The problem Hamiltonian has the form $\displaystyle \sum^n_{i=1} h_i \sigma^i_z + \sum^n_{i=1}\sum^n_{j=1} J_{i,j} \sigma^i_z \sigma^j_z$, where $\sigma^i_z$ is the spin-operator (Pauli-Z operator) for the $i^{th}$ spin, and $h_i$ is the magnetic force acting on the $i^{th}$ spin of the qubit, and $J_{i,j}$ is coupling strengthen between spins $i$ and $j$ of the respective qubits. The spin values are given by $1$ and $-1$, and the final solution of this system, with a high-probability, will be ground state, or state with the lowest energy. Expanding out the terms in a QUBO and defining a map such that $0 \to -1$ and $1 \to 1$, one may then formulate a QUBO in terms of the problem Hamiltonian $H_p$, where the ground state is the solution of the QUBO. This mapping is $1$-to-$1$, ergo, QUBO$\longleftrightarrow$Hamiltonian. See Yarkoni et al. \cite{yarkoni2022quantum} for a more in-depth explanation.  

Therefore, the formulation of quantum annealing has quite a natural mapping to mathematical optimization problems, especially those of combinatorial optimization. Many of these problems are known to be NP-hard \cite{robinson2013introduction}. Several difficult problems, such as the Traveling Salesman Problem, have been known for decades. Surprisingly, there have been applications outside of the well-known operations research problems, such neural networks which may be slightly reformulated as a QUBO optimization problem \cite{sasdelli2021quantum} to solve for the parameters. 

In this work we utilize QA, specifically the hybrid solvers offered by D-Wave, to solve the large combinatorial optimization problem of finding the best subset of features out of many possibilities.  As the feature space grows, the computational complexity of finding the global minima of this combinatorial problem scales exponentially, requiring alternative technology like QA to arrive at a good solution in a reasonable amount of time.  

\section{Methods}
\subsection{Data, Correlation, and Pseudo-Distance Metric}\label{subsec:mathy}
Define the data set as $\big\{(x_i,y_i) \big\}_{i=1}^{N}$, where $x_i\in\mathbb{R}^{n}$ and $y_i\in\mathbb{R}^m$ for all $i$. We call $\mathbf{y} := \{y_i\}_{i=1}^{N}$ the \textit{target variable}, $\mathbf{x} := \{x_i\}_{i=1}^{N}$ \textit{dependent variables}, and for the vectors of the dependent variables, $x_i = (x_i^1, x_i^2, \ldots, x_i^n)$, the set $\mathbf{x}_i :=\{x_i^j\}_{i=1}^{N}$ for each $j=1,\ldots, n$ is called a \textit{feature}.

Given a record of data, denote $I_{i} := I(\mathbf{x}_{i}; \mathbf{y})$ is a scalar that captures the amount of information shared between feature $\mathbf{x}_i$ and the target variable. Similarly, denote $R_{ij} := I(\mathbf{x}_{i}, \mathbf{x}_{j})$ as a scalar that describes the amount of information shared between feature $\mathbf{x}_i$ and feature $\mathbf{x}_j$. While the statistical method in $I_{i}$ and $R_{ij}$ are the same, the difference in notation is given to delineate the linear terms and quadratic terms in the QUBO derived in Subsection \ref{subsec:qubo}.

This statistic in general can be Spearman correlation, mutual information, chi-squared statistic, ANOVA F-test, or any other statistical measure appropriate for the given data set (see \cite{box1978statistics} for further information). The methods applied in this analysis to the quantum feature selection are Spearman correlation and mutual information, and the methods applied to the classical methods are chi-squared statistic and ANOVA F-test.

Spearman correlation is a rank order correlation statistic that enables the comparison of continuous or discrete random variables, generalizing the well-known Pearson correlation \cite{mukaka2012guide}. Denote $R(\mathbf{x})$ as the random variable $\mathbf{x}$ converted to rank, $\mbox{cov}(\cdot, \cdot)$ as the covariance, and $\sigma$ as the standard deviation, then the Spearman coefficient is
\begin{equation}\label{eq:spr}
    r_s = \frac{ \mbox{cov}(R(\mathbf{x}), R(\mathbf{y})) }{ \sigma_{R(\mathbf{x})} \sigma_{R(\mathbf{y})} }.
\end{equation}

Mutual information (MI) is an information theoretic pseudo-metric that leverages probability measures to calculate the shared information by calculating the dependency. The random variables could be continuous or discrete valued. For probability measure $P$, MI has the mathematical form
\begin{equation}\label{eq:mi}
    \mbox{MI}( \mathbf{x}, \mathbf{y}) = \sum_{x,y} P(x,y) \ln \left( \frac{ P(x,y) }{ P(x) P(y) } \right).
\end{equation}

The chi-squared statistic is defined as 
\begin{equation}\label{eq:chi}
    \chi^2 = \sum_{i} \frac{(O_i - E_i)^2}{E_i}
\end{equation}
where $E_i$ is the expected number and $O_i$ is the observed number. 
Chi-squared statistic was derived for discrete random variables. However, if the target is discrete then applications with continuous valued features is possible, such as Pearson correlation can be applied to the discrete random variable of years old and extract a meaningful correlation statistic.  

\begin{figure*}[t!]
    \centering
    \subfigure[]{\includegraphics[width=0.48\textwidth]{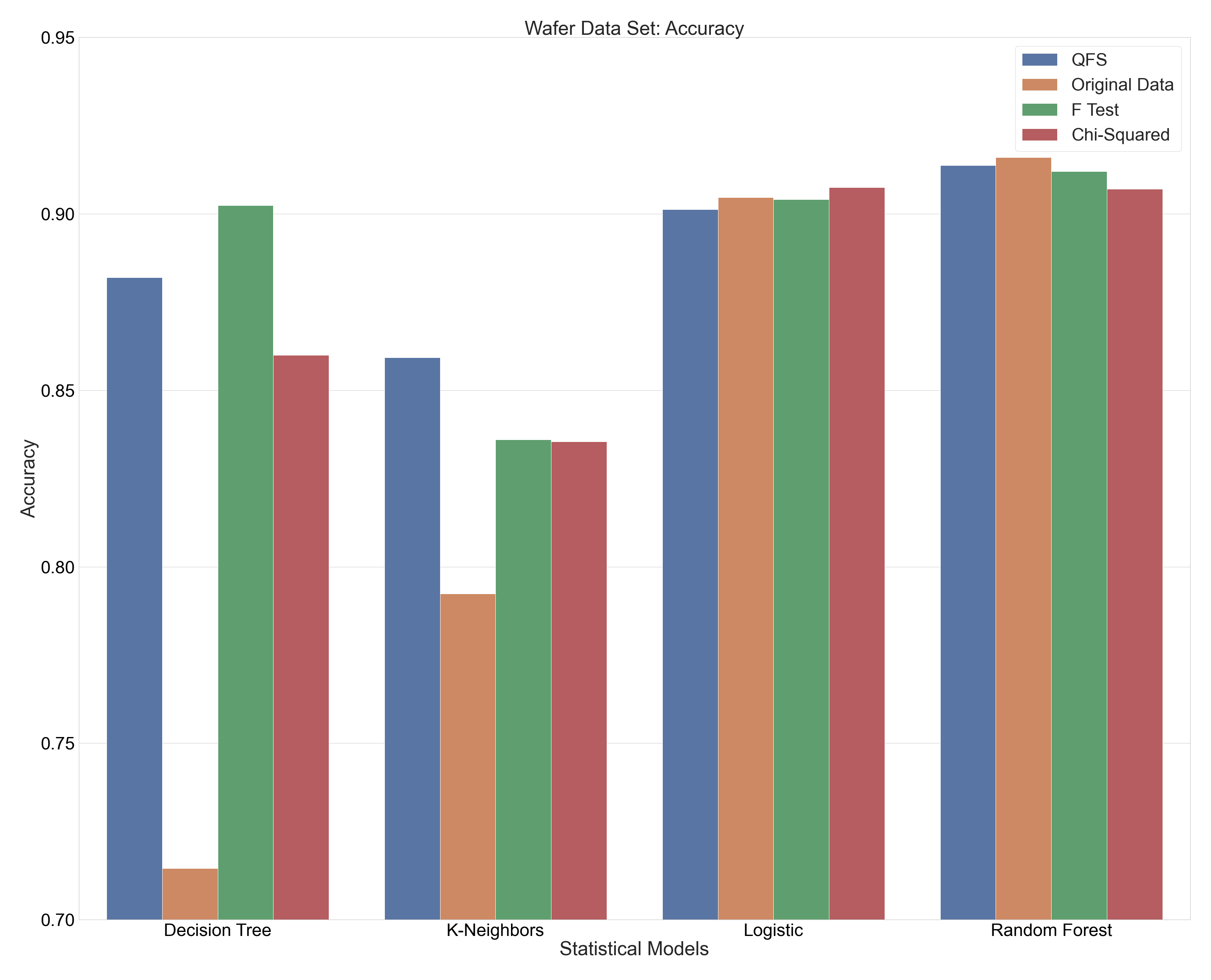}} 
    \subfigure[]{\includegraphics[width=0.48\textwidth]{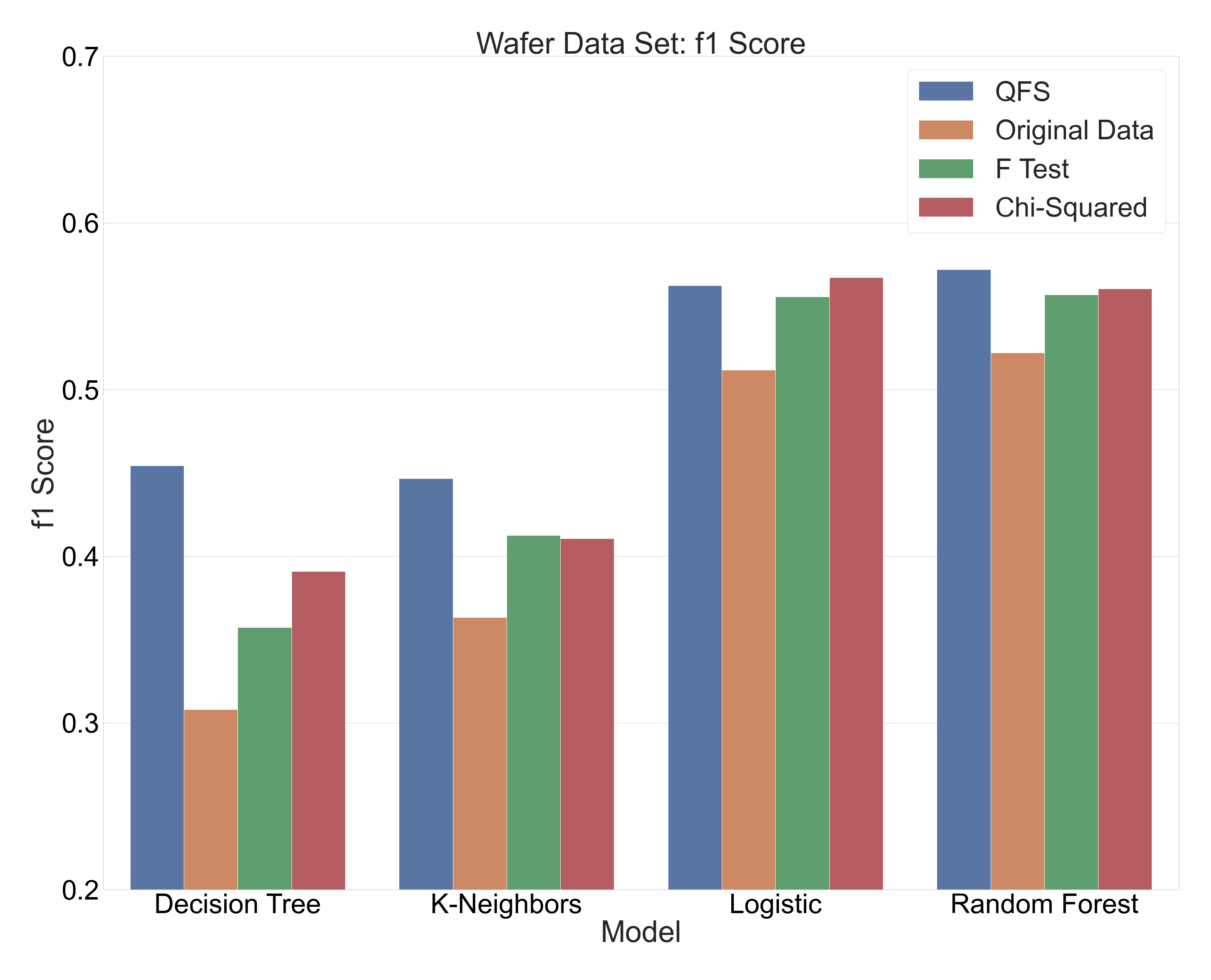}} 
    \caption{Analysis of the Wafer data set where (a) is the accuracy of each classifier and (b) is the $f1$ score }
    \label{fig:wafer}
\end{figure*}

Finally, ANOVA F-test is a means to assess whether the expected values of the general defined groups are different from another. The mathematical formulation is as follows,
\begin{equation}\label{eq:anv}
    f_t = \sum_{i=1}^{K}\sum_{j=1}^{n_i} \frac{ (Y_{ij} - \bar{Y}_i)^2 }{N-K}
\end{equation}
where $\bar{Y}_i$ is the mean in the $i^{th}$-group, $Y_{ij}$ is the $j^th$ observation in the $i^{th}$-group, $K$ denotes the number of groups, and $N$ is the size of the data record.

\subsection{QUBO Derivation}\label{subsec:qubo}
Quantum Feature Selection can be formulated as a combinatorial optimization problem, specifically as a QUBO. For the QUBO, denote $b_{i}$ as a binary variable that is $1$ if feature $i$ is selected for inclusion and $0$ otherwise. Now, defining $\mathbf{b} = (b_1,\ldots,b_n)$ and $||\mathbf{b}||_1 = \sum_{i=1}^{n} b_i$, for a given number of features, say $k$, the QUBO has the form

\begin{equation}\label{eq:formu}
    \argmin_{ \mathbf{b}, \ ||\mathbf{b}||_1 = k } = -\alpha \cdot \sum_{i=1}^{n}I_{i}b_{i} + \big( 1-\alpha \big) \cdot \vspace{-5pt} \sum_{i,j=1 ,  \ i\neq j}^{n} R_{ij}b_{i}b_{j}.
\end{equation}

We follow the technique in \cite{mucke2022quantum} where we utilize a binary search algorithm to find the optimal value of $\alpha$ to arrive at the targeted number of features.  $\alpha$ is an intuitive and natural way to control the number of features: a value of close to $1$ will select all features with some level of predictive power, and a value close to $0$ will only select the small number of features that possess completely unique amounts of information not found in any other feature.  While the optimal value of alpha will differ depending on the data set and information metric used, the binary search algorithm outlined in \cite{mucke2022quantum} provides a clear and efficient way for attaining that value with a small number of iterations.

However, given the number of iterations required for this algorithm, utilization of a quantum annealer processor, such as the processor developed by D-wave \cite{Dwave}, has allowed this algorithm to be efficiently run in a timely manner with respect to the number of features. 

\begin{figure*}[t!]
    \centering
    \subfigure[]{\includegraphics[width=0.48\textwidth]{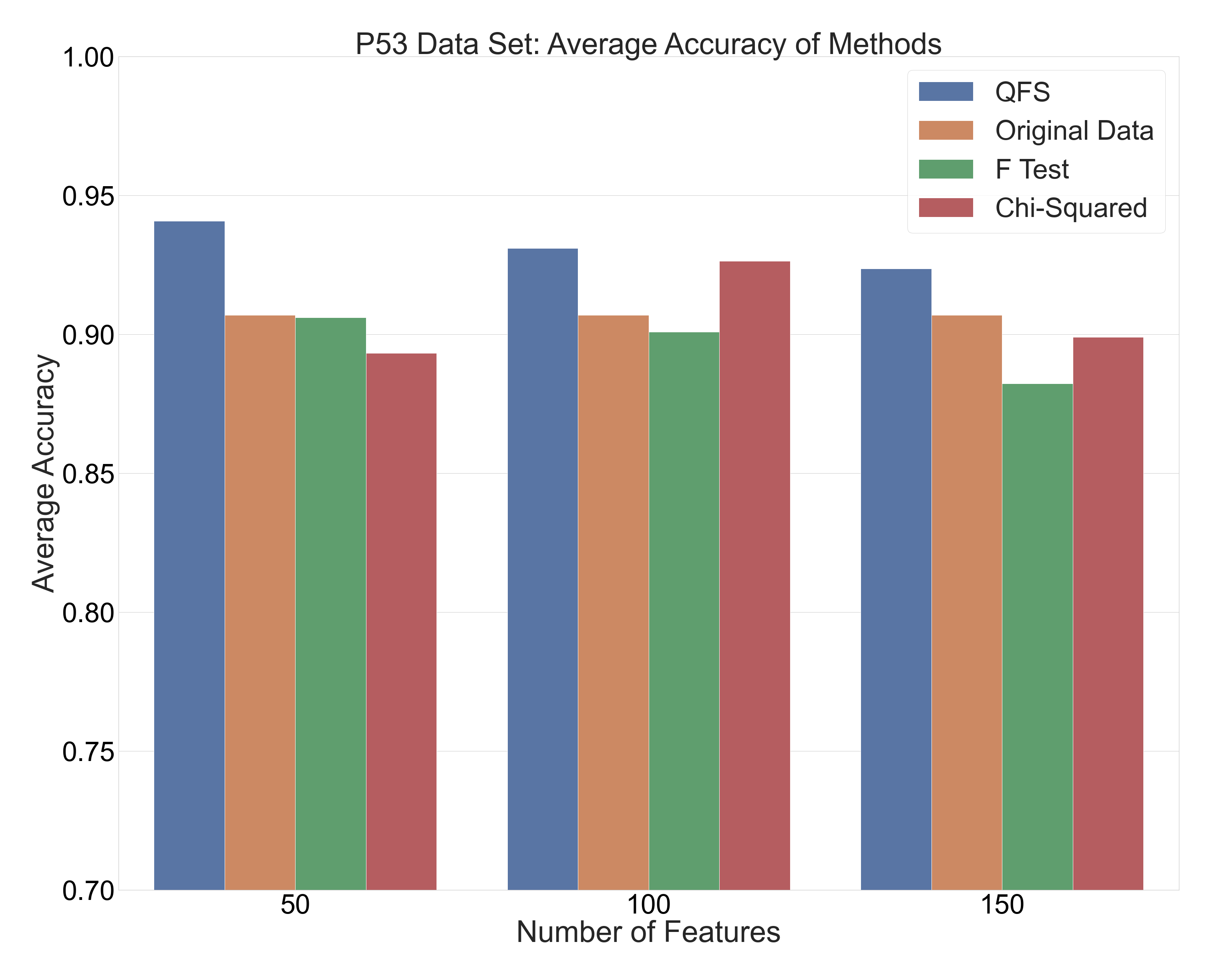}} 
    \subfigure[]{\includegraphics[width=0.48\textwidth]{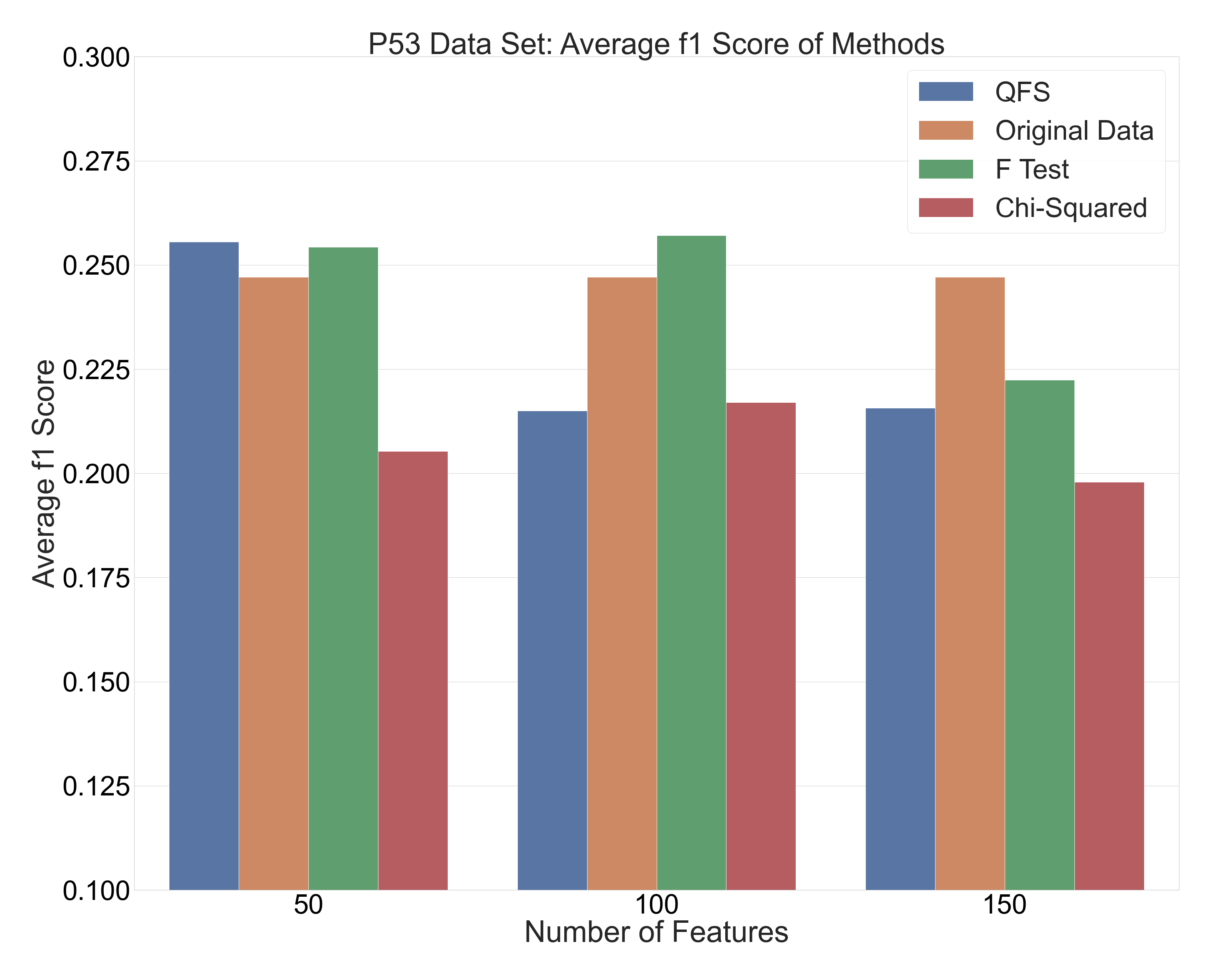}}
    \caption{Analysis of the P53 data set where (a) is the average accuracy of each method with respect to the number of features, and (b) is the average $f1$ score of each method with respect to the number of features. Increments of 50 were selected to sample progression of the models. 
    }
    \label{fig:p53}
\end{figure*}

\subsection{Data and Experimentation}
The data sets applied in the analysis were the p53 mutants data set (denoted as \textit{P53}) and the relative location of CT slices on axial axis data set (dented as \textit{CT}) from the UCI Machine Learning Repository \cite{Dua:2019}, and the data set for the Detecting Anomalies in Wafer Manufacturing\footnote{\href{https://www.kaggle.com/datasets/arbazkhan971/anomaly-detection}{ https://www.kaggle.com/datasets/arbazkhan971/anomaly-detection}} (denoted as \textit{Wafer}) Kaggle competitions. Note that the heterogeneity of the data sets where selected to sample how well QFS performed in general. 

As noted in the Subsection \ref{subsec:mathy}, the ANOVA F-test and chi-squared statistic are the two classical features selection algorithms. For a general basis for comparison, the entire data set is also applied to each model. Features selected by the quantum feature selection algorithm are denoted by \textit{QFS}, features selected by ANOVA F-test are denoted by \textit{F Test}, and features selected by the chi-squared statistic is denoted by \textit{Chi-Squared}. When the entire record of data is applied it is denoted by \textit{Original Data}. 

For robust testing, the following classification models are applied: decision trees, random forests, logistic regression, and k-nearest neighbors, see \cite{bonaccorso2017machine} for more information. The classifiers selected for the analysis sample techniques from clustering, weak learners, and kernel-like methods. For regression, decision tree regressor, random forest regressor, Adaboost regressor, and k-nearest neighbors regressor algorithms were applied. For the modeling analysis a $5$-fold sampling of the data is taken and the results are averaged for each respective algorithm. Note that state-of-the-art algorithms require parameter tuning, which has a tremendous effect on model performance. To bypass such questions, the application of simpler, yet robust, algorithms were taken for the initial experiments.  

All algorithms applied in the analysis, including preprocessing the data, were written in Python and taken from the Scikit-Learn library \cite{pedregosa2011scikit}.

\section{Numerical Analysis}

The first data set considered is Wafer. There are 1763 data points, 1558 features, and 143 data points identifying a defect. This is an approximate ratio of 11:1 with the classes. The features are a mixture of continuous and binary variables. The data was not pre-processed, aside from randomly sampling the data and splitting it into the train, validation and test sets.  With this ratio of the classes a low $f1$ score is expected. Adjustments for the class imbalance was not implemented before training the statistical model in order to not incorporate unintended bias. Lastly, mutual information was utilized as the statistical measure for QFS. 

\begin{figure*}[!htbp]
    \centering
    \subfigure[]{\includegraphics[width=0.48\textwidth]{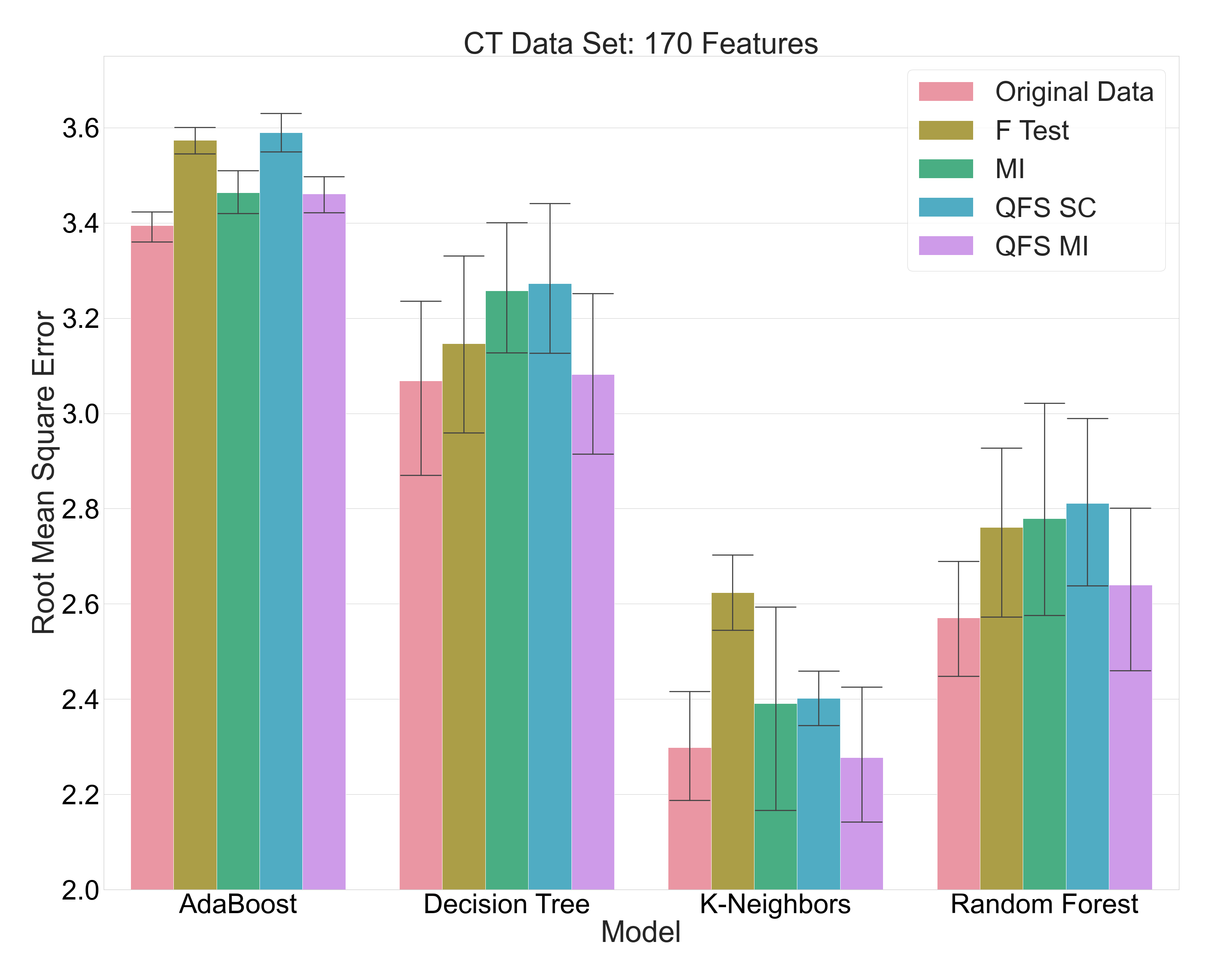}} 
    \subfigure[]{\includegraphics[width=0.48\textwidth]{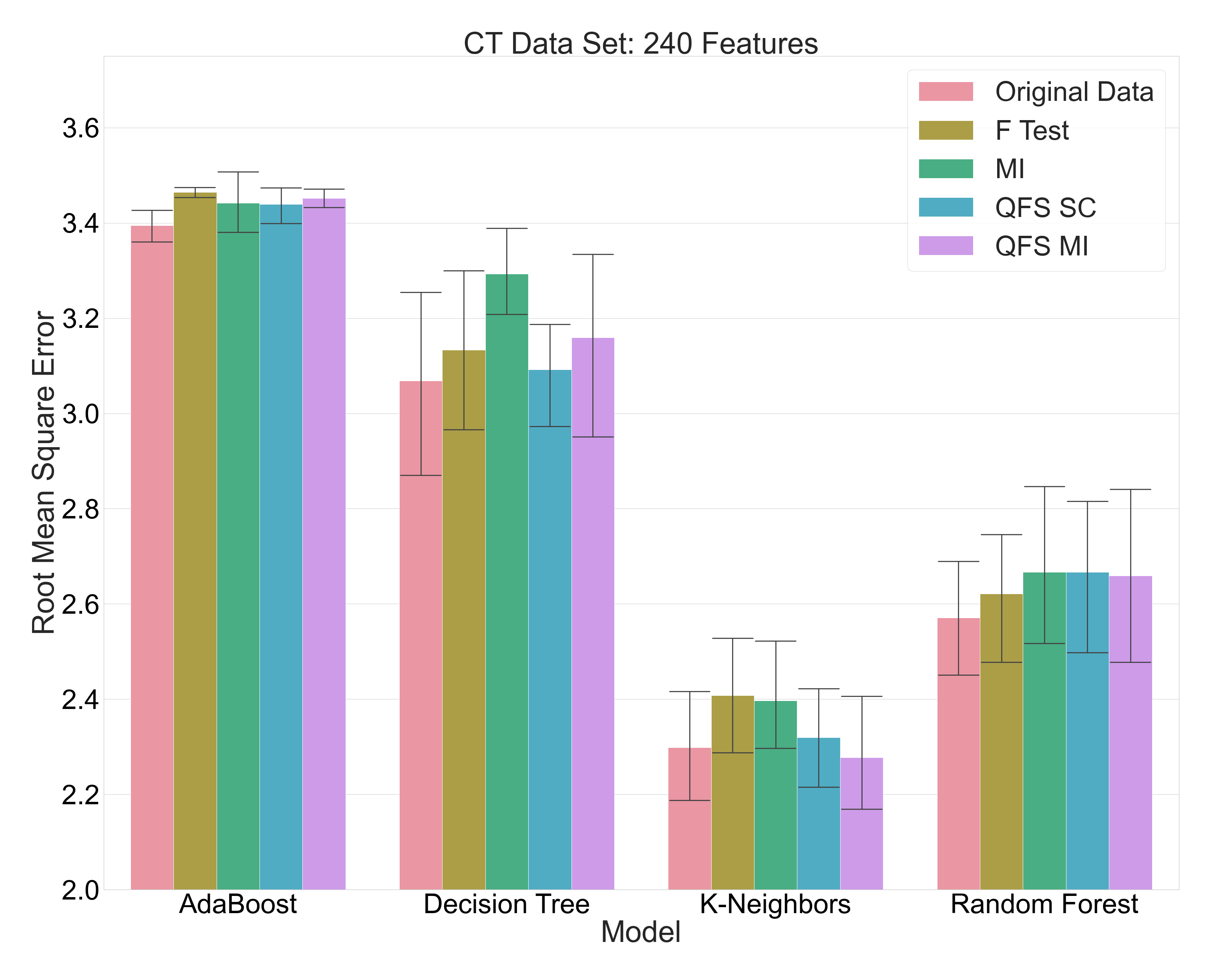}}
    \caption{Analysis of the CT data set where (a) average root mean square error and error bar with 170 features selected, and (b) average root mean square error and error error bar with 240 features selected. The feature of 170 and 240 were selected after initial expirements. 
    }
    \label{fig:ct}
\end{figure*}

For this analysis 100 features were identified from each feature selection method. Aggregating the results of model performance, QFS had a slight advantage over the classical methods, and for model stability QFS had a clear advantage. For more granular analysis Figure \ref{fig:wafer} displays the outcomes for each classification algorithm.

For the P53 data set the analysis is extended to include 50 features, 100 features, and 150 features. The extended analysis was motivated by the number of features of the data, which are a total of 5408 features. The P53 data set had a total of 31,420 data points with 151 identified instances of the mutation, giving the approximate class ratio of 205:1. The features are a mixture of continuous and binary variables. 

\begin{table}[!htbp]
{ \begin{tabular}{ |c|c|c|} \hline
Method & Average Accuracy & Average $f1$ Score \\ \hline
QFS & \textcolor{red}{0.889} & \textcolor{red}{0.509}\\ \hline
F Test & 0.889 & 0.471\\ \hline
Chi-Squared &  0.877 & 0.482 \\ \hline
Original Data & 0.832 & 0.426 \\ \hline
\end{tabular} }
 \caption{Aggregated results of the Wafer data set, where the best scores are highlighted in red.}\label{tab:compare}
 \end{table}

Following the same process as the Wafer data set, including using mutual information as the statistical measure, the aggregated results display an advantage with the QFS technique when 50 features are selected; the results are aggregated with number of features are displayed in Figure \ref{fig:p53}. However, the features selected by the QFS algorithm do not perform as well as the classical methods. Interestingly, there is a noticeable decrease in the average model performance and average model stability from each respective increase in the number of features, not including the classical methods average $f1$ score from 50 to 100 hundred features and the average $f1$ score of QFS from 100 to 150 features, which remains the same. While there is an increase in model stability from 50 to 100 features for the classical methods, there is a significant decrease from 100 to 150 features. These results indicate noise in the data and hint at an optimal number of features to select. Moreover, the lack of model stability from the features selected by QFS could stem from the mutual information statistic that focuses more on the majority class, and other statistical methods may increase the discovering the signal from the noise. Finally, from the stability of the $f1$ scores for QFS from 100 to 150 features it appears that algorithm did pick up on the noise but then minimized the impact.

Lastly, for the CT data set, 170 features and 240 features were selected. this data set has 384 features, 53,500 data points, where the features are a mix of continuous and nominal. The target variable is continuous.

Given the more complicated task of regression, for the classical feature selection algorithms ANOVA F-test and mutual information were chosen. Just as before, ANOVA F-test is denoted as \textit{F-test} and mutual information is denoted as \textit{MI}. For QFS, mutual information and Spearman rank correlation were independently applied to obtain statistical information and QFS was ran to compare both techniques. These are denoted as \textit{QFS MI} and \textit{QFS SC}, respectively.   

Figure \ref{fig:ct} displays the average root mean square error for 170 features and 240 features selected, respectively. Except for k-neighbors regressor algorithm, unlike the classification statistical modeling, the entire data set out-performed each feature selection method. However, with the k-neighbor regressor the QFS MI selection algorithm not only out-performed the original data, (although, not by much), this regression model out-performed all of models. Moreover, the regression models trained by the 170 features selected by QFS out-performed all of the other respective regression models trained with the other selected 170 features, as well as be competitive with the performance of the regression models trained with all of the data. 

The the algorithms for the regression were identified a priori and chosen for their simplicity. Given that the k-neighbor regressor performed the best, as well as poor performance for all statistical models, this suggested more complexity algorithms to apply the regression is required for higher performing and more robust predictor.

Interestingly, the performance of the regression models trained with 240 selected with the QFS MI algorithm drops. Moreover, for each training algorithm with 240 features selected, excluding the models trained with all of the data, the top performing respective regression model is not clear. For instance, the 170 features selected by the QFS SC algorithm performs poorly across all training algorithms, and in contrary, with 240 features QFS SC is competitive with all of the data applying the decision tree regression algorithm. This indicates there is a lot more noise in the data which appears to average out when all of the data is utilized for training.

\section{Discussion} Through numerical analysis it was demonstrated that, on average, the features select by the QFS algorithm enabled the training of models more stable than the features selected from the classical techniques. Moreover, the results demonstrate the need to consider other statistical techniques, including mixing techniques and averaging techniques contingent on the data quality and feature type, to assist further separating the signal from the noise. However, mutual information was numerically demonstrated to do well as a statistical measure of shared information.    

Suggested further research includes the utilization of a larger temporal stamped data set to test how each model would perform with the temporal out-of-sample data, mimicking a model in production, and the careful application of state-of-the-art algorithms, ensuring hyperparameter tuning does not have an effect on the performance of the model.   

\section{Disclaimer}
About Deloitte: Deloitte refers to one or more of Deloitte Touche Tohmatsu Limited (“DTTL”), its global network of member firms, and their related entities (collectively, the “Deloitte organization”). DTTL (also referred to as “Deloitte Global”) and each of its member firms and related entities are legally separate and independent entities, which cannot obligate or bind each other in respect of third parties. DTTL and each DTTL member firm and related entity is liable only for its own acts and omissions, and not those of each other. DTTL does not provide services to clients. Please see www.deloitte.com/about to learn more.

Deloitte is a leading global provider of audit and assurance, consulting, financial advisory, risk advisory, tax and related services. Our global network of member firms and related entities in more than 150 countries and territories (collectively, the “Deloitte organization”) serves four out of five Fortune Global 500® companies. Learn how Deloitte’s
approximately 330,000 people make an impact that matters at www.deloitte.com. 
This communication contains general information only, and none of Deloitte Touche Tohmatsu Limited (“DTTL”), its global network of member firms or their related entities (collectively, the “Deloitte organization”) is, by means of this communication, rendering professional advice or services. Before making any decision or taking any action that
may affect your finances or your business, you should consult a qualified professional adviser. No representations, warranties or undertakings (express or implied) are given as to the accuracy or completeness of the information in this communication, and none of DTTL, its member firms, related entities, employees or agents shall be liable or
responsible for any loss or damage whatsoever arising directly or indirectly in connection with any person relying on this communication. 
Copyright © 2022. For information contact Deloitte Global.

\bibliographystyle{unsrt}
\bibliography{fsbib}

\end{document}